\documentclass[prl, groupedaddress, showkeys,showpacs]{revtex4}

\usepackage{graphicx}
\begin{document}


\title{Current in a three-dimensional periodic tube with unbiased forces}


\author{Bao-quan  Ai$^{a}$}\email[Email: ]{aibq@scnu.edu.cn}
\author{Liang-gang Liu$^{b}$}

\affiliation{$^{a}$ School of Physics and Telecommunication
Engineering, South China
Normal University, 510006 GuangZhou, China.\\
$^{b}$ The Faculty of Information Technology , Macau University of
Science and Technology, Macao.}


\date{\today}
\begin{abstract}
\baselineskip 0.2in \indent Transport of a Brownian particle
moving along the axis of a three-dimensional asymmetric periodic
tube is investigated in the presence of asymmetric unbiased
forces. The reduction of the coordinates may involve not only the
appearance of entropic barrier but also the effective diffusion
coefficient. It is found that in the presence of entropic barrier,
the asymmetry of the tube shape and the asymmetry of the unbiased
forces are the two ways of inducing a net current. The current is
a peaked function of temperature which indicates that the thermal
noise may facilitate the transport even in the presence of
entropic barrier. There exists an optimized radius at the
bottleneck at which the current takes its maximum value.
Competition between the two opposite driving factors may induce
current reversal.
\end{abstract}

\pacs{05. 60. Cd, 05. 40. Jc, 02. 50. Ey }
\keywords{Entropic barrier, Brownian motors, Current}



\maketitle

\section {Introduction}
\indent Transport phenomena play a crucial role in many processes
from physical, biological to social systems. There has been an
increasing interest in transport properties of nonlinear systems
which can extract usable work from unbiased nonequilibrium
fluctuations \cite{1,2,3,4,5}. In these systems, directed Brownian
motion of particles is generated by nonequilibrium noise in the
absence of any net macroscopic forces and potential gradients
\cite{4}. In all these studies, the potential is taken to be
asymmetric in space. It has also been shown that a unidirectional
current can also appear for spatially symmetric potentials if
there exits an external random force either asymmetric \cite{2} or
spatially-dependent \cite{3}.

\indent The most studies have been referred to the consideration
of the energy barrier. The nature of the barrier depends on which
thermodynamic potential varies when passing from one well to the
other, and their presence plays an important role in the dynamics
of the solid state physics system. However, in some cases, such as
soft condensed matter and biological systems, the entropy barriers
should be considered. The entropy barriers may appear when
coarsening the description of a complex system  for simplifying
its dynamics. Reguera and co-workers \cite{6} use the mesoscopic
nonequilibrium thermodynamics theory
 to derive the general kinetic equation of a system and analyze in
 detail the case of diffusion in a domain of irregular geometry in
 which the presence of the boundaries induces an entropy barrier
 when approaching the dynamics by a coarsening of the description.
 In their recent work \cite{7}, they study the current and diffusion of a
 Brownian particle in a symmetric channel with a biased external
 force. They found that the temperature dictates the strength of the
 entorpic potential, and thus an increasing of temperature leads to
 a reduction of the current. The previous
 works on entropic barriers is limited to biased external force.  However, in some cases, the net current
 occurs in the absence of any net macroscopic forces or in the presence of unbiased forces. For example,
 Molecular motors move along the microtubule without any net
 macroscopic force.
 The present work is extend the study of entropic
 barriers to case of unbiased forces that indicates zero
 force at macroscopic scale. Our emphasis is on finding conditions of obtaining a net current in
 the presence of entropic barriers. The asymmetry of the tube shape and the asymmetry of the unbiased
 fluctuations are the two driving factors for obtaining a net current.
 When the two driving factors compete each other, the current may reverse its
 direction.

\section {Current in a three-dimensional periodic tube}
\begin{figure}[htbp]
  \begin{center}\includegraphics[width=10cm,height=6cm]{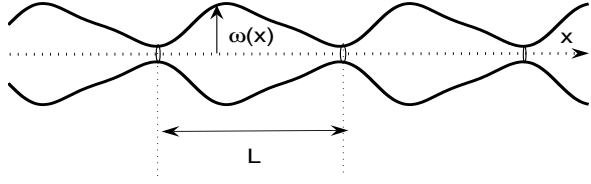}
  \caption{ \baselineskip 0.4in Schematic diagram of a tube with periodicity $L$. The
  shape is described by the radius of the tube $ \omega(x)=a[\sin(\frac{2\pi x}{L})+\frac{\Delta}{4}\sin(\frac{4\pi
    x}{L})]+b$. $\Delta$ is the asymmetric parameter of the tube shape.}\label{1}
\end{center}
\end{figure}

\indent We consider a Brownian particle moving in a asymmetric
periodic tube [Fig. 1] in the presence of an asymmetric unbiased
fluctuations. Its over-damped dynamics is described by \cite{7}
\begin{equation}\label{}
    \eta\frac{d \vec{r}}{dt}=\vec{F}_{x}(t)+\sqrt{\eta
    k_{B}T}\vec{\xi}(t),
\end{equation}
where $\vec{r}$ is the 3D coordinate, $\eta$ the friction
coefficient of the particle, $k_{B}$ the Boltzmann constant, $T$
the temperature, and $\overrightarrow{\xi}(t)$ is Gaussian whit
noise with zero mean and correlation function:
$<\xi_{i}(t)\xi_{j}(t^{'})>=2\delta_{i,j}\delta(t-t^{'})$ for
$i,j=x, y, z$. $<...>$ denotes an ensemble average over the
distribution of $\overrightarrow{\xi}(t)$. $\delta(t)$ is the
Dirac delta function. The reflecting boundary conditions ensure
the confinement of the dynamics with the tube. $\vec{F}_{x}(t)$ is
an asymmetric unbiased external force along $x$ direction. $F(t)$
is its scalar quantity and satisfies\cite{8,9}
\begin{equation}
F(t+\tau)=F(t), \int^{\tau}_{0}F(t)dt=0,
\end{equation}
\begin{equation}\label{}
 F(t)=\left\{
\begin{array}{ll}
   \frac{1+\varepsilon}{1-\varepsilon}F_{0},& \hbox{$n\tau\leq
t<n\tau+\frac{1}{2}\tau(1-\varepsilon)$};\\
   -F_{0} ,&\hbox{$n\tau+\frac{1}{2}\tau(1-\varepsilon)<t\leq
      (n+1)\tau$},\\
\end{array}
\right.
\end{equation}
where $\tau$ is the period of the unbiased force , $F_{0}$ its
magnitude and $\varepsilon$ the temporal asymmetric parameter with
$-1\leq \varepsilon \leq 1$.

 \indent The shape of the tube is described by its radius
\begin{equation}\label{}
    \omega(x)=a[\sin(\frac{2\pi x}{L})+\frac{\Delta}{4}\sin(\frac{4\pi
    x}{L})]+b,
\end{equation}
where $a$ is the parameter that controls the slope of the tube,
$\Delta$ the asymmetry parameter of the tube shape. The radius at
the bottleneck is $r_{b}=b-a\sqrt{1+\frac{\Delta^{2}}{16}}$.

\indent The movement equation of a Brownian particle moving along
the axis of the 3D tube can be described by the Fick-Jacobs
equation \cite{6,7,10,11} which is  derived from the 3D (or 2D)
Smoluchowski equation after elimination of $y$ and $z$ coordinates
by assuming equilibrium in the orthogonal directions. The
reduction of the coordinates may involve not only the appearance
of entropic barrier, but also the effective diffusion coefficient.
When $|\omega^{'}(x)|<<1$, the effective diffusion coefficient
reads \cite{7,10}
\begin{equation}\label{}
    D(x)=\frac{D_{0}}{[1+\omega^{'}(x)^{2}]^{\alpha}},
\end{equation}
where $D_{0}=k_{B}T/\eta$ and $\alpha=1/3$, $1/2$ for 2D and 3D,
respectively.

\indent Consider the effective diffusion coefficient and the
entropic barrier, the dynamics of a Brownian particle moving along
the axis of the 3D tube can be described by \cite{6,7}
\begin{equation}\label{}
    \frac{\partial P(x,t)}{\partial t}=\frac{\partial}{\partial x}[D(x)\frac{\partial P(x,t)}{\partial
    x}+\frac{D(x)}{k_{B}T}\frac{\partial A(x,t)}{\partial x}P(x,t)]=-\frac{\partial j(x,t)}{\partial
    x},
\end{equation}
where we defines a free energy $A(x,t):=E-TS=-F(t)x-Tk_{B}\ln
h(x)$, here $E=-F(t)x$ is the energy, $S=k_{B}\ln h(x)$ the
entropy, $h(x)$ the dimensionless width $2\omega(x)/L$ in 2D, and
the dimensionless transverse cross section $\pi[\omega(x)/L]^{2}$
of the tube in 3D. $j(x,t)$ is the probability current density.
$P(x,t)$ is the probability density for the particle at position
$x$ and at time $t$. It satisfies the normalization condition
$\int_{0}^{L}P(x,t)dx=1$ and the periodicity condition
$P(x,t)=P(x+L,t)$.

\indent If $F(t)$ changes very slowly with respect to $t$, namely,
its period is longer than any other time scale of the system,
 there exists a quasi-steady state. In this case, by following the  method in  \cite{1,2,3,4,5,6,7,12}, we can obtain the
current
\begin{equation}\label{}
    j(F(t))=\frac{k_{B}T[1-\exp(-\frac{F(t)L}{k_{B}T})]}{\int_{0}^{L}h(x)\exp(\frac{F(t)x}{k_{B}T})dx\int_{x}^{x+L}[1+\omega^{'}(y)^{2}]^{\alpha}h^{-1}(y)\exp(-\frac{F(t)y}{k_{B}T})dy}.
\end{equation}

\indent The average current is
\begin{equation}\label{9}
    J=\frac{1}{\tau}\int_{0}^{\tau}j(F(t))dt=\frac{1}{2}(j_{1}+j_{2}),
\end{equation}
with
\begin{equation}\label{}
    j_{1}=(1-\varepsilon)j(\frac{1+\varepsilon}{1-\varepsilon}F_{0}),\indent
    j_{2}=(1+\varepsilon)j(-F_{0}).
\end{equation}

\section {Results and discussions}

 \indent  Because the results  from 2D and 3D are very similar, for
the convenience of physical discussion, we now mainly investigated
the current in 3D  with $k_{B}=1$ and $\eta=1$.

\begin{figure}[htbp]
  \begin{center}\includegraphics[width=10cm,height=8cm]{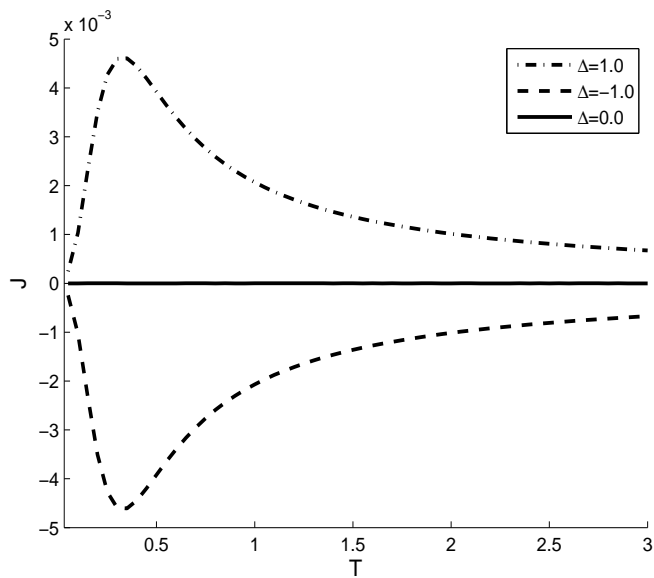}
  \caption{\baselineskip 0.4in Current $J$ versus temperature $T$ for different asymmetric parameters $\Delta$ of the tube shape at $a=1/2\pi$, $b=1.5/2\pi$, $L=2\pi$, $\alpha=1/2$, $F_{0}=0.5$ and $\varepsilon=0.0$.}\label{1}
\end{center}
\end{figure}

\indent The current $J$ as a function of temperature $T$ for the
case $\varepsilon=0.0$ is presented in Fig. 2 for different values
of asymmetric parameter of the tube shape. The curve is observed
to be bell shaped, which shows the feature of resonance. When
$T\rightarrow 0$, the particle cannot reach $3D$ area and the
effect of entropic barrier disappears and there is no current.
When $T\rightarrow \infty $, the effect of the unbiased forces
disappears and the current goes to zero, also. There is an
optimized value of $T$ at which the current $J$ takes its maximum
value, which indicates that the thermal noise may facilitate the
particle transport even in the presence of entropic barrier. The
current is negative for $\Delta<0$, zero at $\Delta=0$, and
positive for $\Delta>0$. Therefore, the asymmetry of the tube
shape is a way of inducing a net current.

\begin{figure}[htbp]
  \begin{center}\includegraphics[width=10cm,height=8cm]{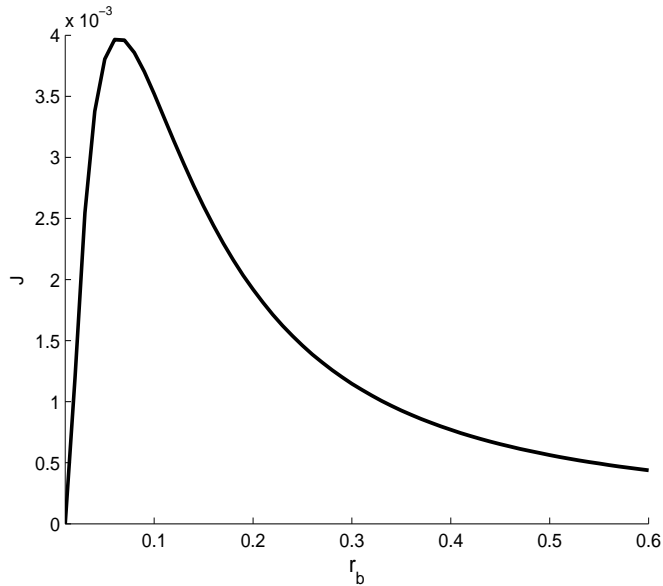}
  \caption{\baselineskip 0.4in Current $J$ versus the radius $r_{b}$ at the bottleneck at $a=1/2\pi$, $\alpha=1/2$, $L=2\pi$, $F_{0}=0.5$, $T=0.5$, $\Delta=1.0$ and $\varepsilon=0.0$.}\label{1}
\end{center}
\end{figure}

\indent Figure 3 shows the current $J$ as a function of the radius
at the bottleneck $r_{b}$ for the case $\varepsilon=0.0$ and
$\Delta=1.0$.  If the bottleneck has zero the particle can not pass
through the bottleneck, so the current should be zero. When the
bottleneck has infinite radius, the effect of tube shape disappears
and the current tends to zero, also. Therefore, the current $J$ is a
peaked function of the radius $r_{b}$ at the bottleneck.

\begin{figure}[htbp]
  \begin{center}\includegraphics[width=10cm,height=8cm]{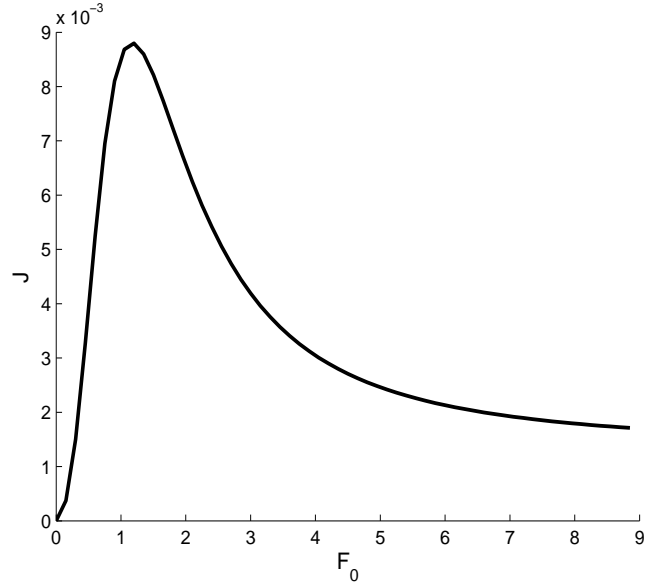}
  \caption{\baselineskip 0.4in Current $J$ versus amplitude $F_{0}$ of the external force at $a=1/2\pi$, $b=1.5/2\pi$, $L=2\pi$, $\alpha=1/2$, $T_{0}=0.5$ and $\varepsilon=0.0$.}\label{1}
\end{center}
\end{figure}

\indent Figure 4 shows the current $J$ versus the amplitude
$F_{0}$ of the unbiased forces of for the case $\varepsilon=0.0$
and $\Delta=1.0$. When $F_{0}\rightarrow 0$, only the effect of
the entropic barrier exists, so the current tends to zero. The
current $J$ saturates to a certain value in large amplitude
$F_{0}$ limit. There exists an optimized value of $F_{0}$ at which
the current $T$ takes its maximum value.

\begin{figure}[htbp]
  \begin{center}\includegraphics[width=10cm,height=8cm]{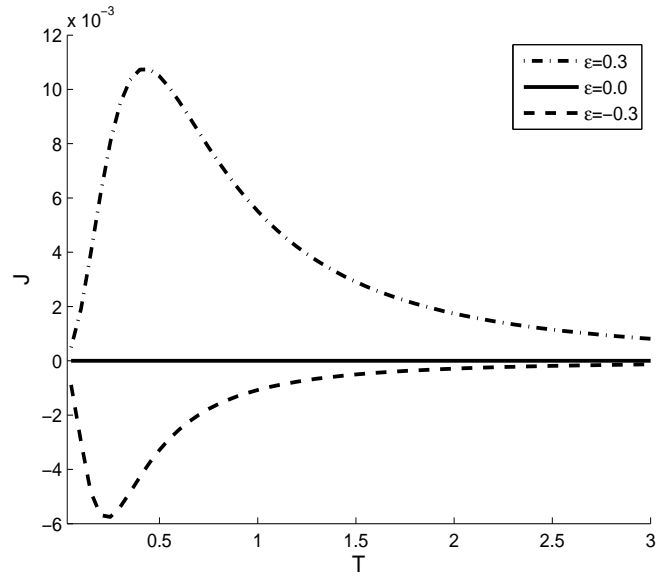}
  \caption{\baselineskip 0.4in Current $J$ versus temperature $T$ for different asymmetric parameters  $\varepsilon$  of the external force at $a=1/2\pi$, $b=1.5/2\pi$, $L=2\pi$, $\alpha=1/2$, $F_{0}=0.5$ and $\Delta=0.0$.}\label{1}
\end{center}
\end{figure}
\indent The current $J$ as a function of temperature $T$ for the
case $\Delta=0.0$ is presented in Fig. 5 for different values of
asymmetric parameter $\varepsilon$ of unbiased forces. The curve
is a bell shaped function of temperature which is the same as that
in Fig. 2. The asymmetry of the unbiased forces can induce a net
current even when the shape of the tube is symmetric. For the
asymmetry of the tube shape, the asymmetry of the external force
is another way of obtaining a net current. Similarly, $J$ is
negative for $\varepsilon<0$, zero at $\varepsilon=0$ and positive
for $\varepsilon>0$.

\begin{figure}[htbp]
  \begin{center}\includegraphics[width=10cm,height=8cm]{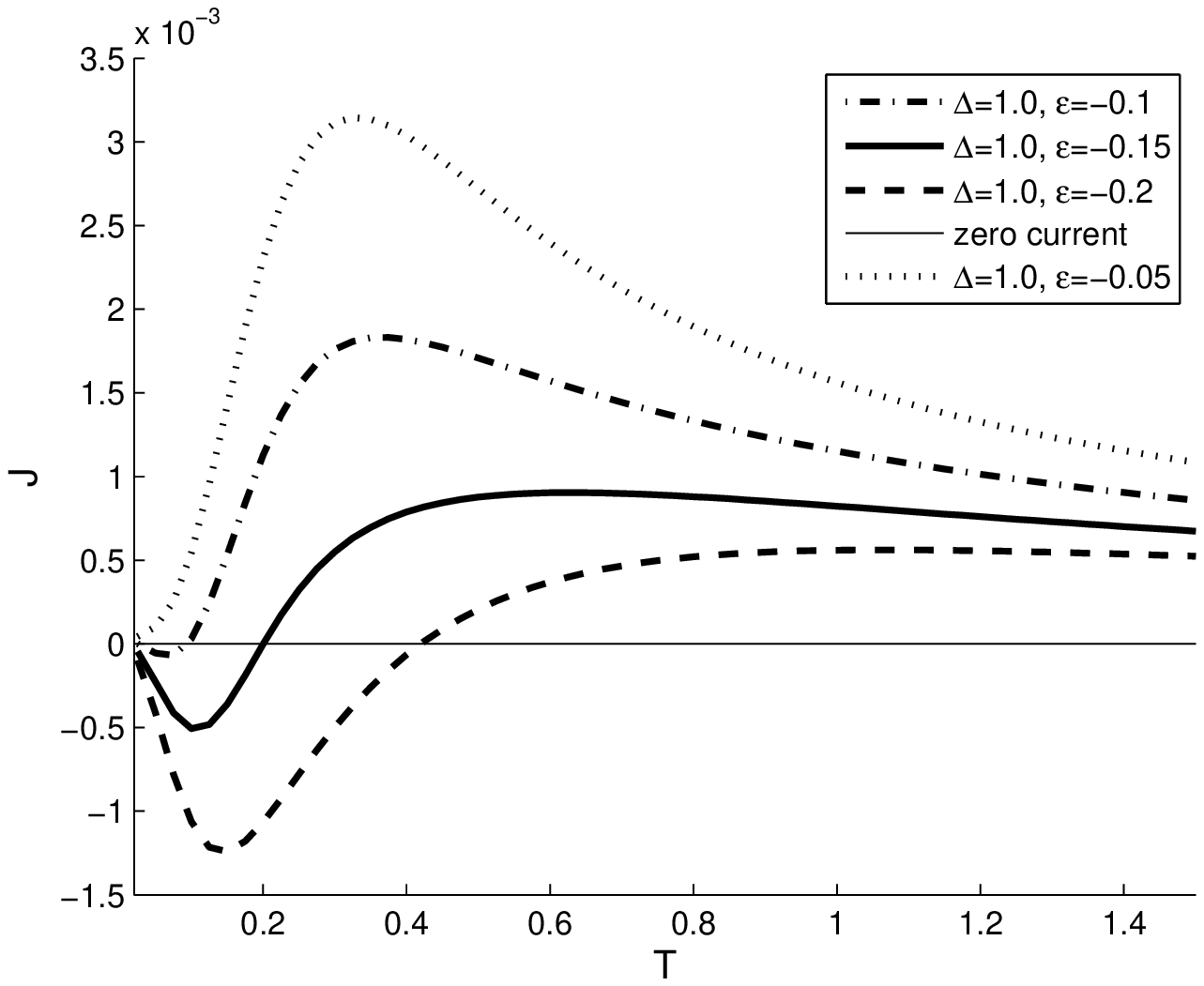}
  \caption{\baselineskip 0.4in Current $J$ versus temperature $T$ for different combinations of $\varepsilon$ and $\Delta$ at $a=1/2\pi$, $b=1.5/2\pi$, $L=2\pi$, $\alpha=1/2$ and $F_{0}=0.5$.}\label{1}
\end{center}
\end{figure}
\indent In Fig. 6, we plot the current $J$ as a function of
temperature $T$ for different combinations of $\Delta$ and
$\varepsilon$. When the asymmetric parameter $\Delta$ of the tube
shape is positive, the current may reversal its direction on
increasing temperature for negative $\varepsilon$. It is obvious
that the current reversal may occur when a positive driving factor
competes with a negative one. Therefore, there may exist current
reversal for $\Delta\varepsilon<0$ ($\Delta=1.0$,
$\varepsilon=-0.1$; $\Delta=1.0$, $\varepsilon=-0.15$;
$\Delta=1.0$, $\varepsilon=-0.2$). However, $\Delta\varepsilon<0$
is not a sufficient condition for current reversal. For example,
the current is always positive for $\Delta=1.0$,
$\varepsilon=-0.05$.

\section{Concluding Remarks}
  \indent In this paper, we study the transport of a Brownian
  particle moving in a 3D periodic tube in the presence of the unbiased forces.
  The presence of the boundaries induces an entropic
  barrier when approaching the exact dynamics by coarsening of the
  description. Both the asymmetry of the tube shape and the asymmetry
  of the unbiased forces are the two driving factors for
  obtaining a net current. When the two driving factors compete
  each other, the current may reverse its direction upon
  increasing temperature. When the bottleneck has zero radius and infinite radius
 then the current tends to zero. An optimized bottleneck radius
 leads to a maximum current. There exists an
  optimize value of temperature at which the current takes its
  maximum value, which indicates that the thermal noise may
  facilitate the current. The results we have presented have a
  wide application in many processes \cite{6}, such as molecular motors
  movement through the microtubule in the absence of any net macroscopic
  forces \cite{13},  ions transport through ion channels \cite{14}, motion of polymers
  subjected to rigid constraints \cite{15}, drug release \cite{16}and polymer
  crystallization \cite{17}. In these systems, a directed-transport with entropic barriers can be
  obtained in the absence of any net macroscopic forces or in the presence of the unbiased forces.

\section{ACKNOWLEDGMENTS}
The authors thank Prof. Masahiro NAKANO for helpful discussions.
The work is supported by the National Natural Science Foundation
of China under Grant No. 30600122 and  GuangDong Provincial
Natural Science Foundation under Grant No. 06025073.

\end{document}